\documentclass[12pt]{article}
\usepackage{epsfig,graphicx}
\usepackage{fpcp04}

\def\beq{\begin{equation}}
\def\eeq#1{\label{#1}\end{equation}}
\def\eeqn{\end{equation}}

\def\beqa{\begin{eqnarray}}
\def\eeqa#1{\label{#1}\end{eqnarray}}
\def\eeqan{\end{eqnarray}}

\let\bar=\overbar

\def\Dslash{\not{\hbox{\kern-4pt $D$}}}
\def\dslash{\not{\hbox{\kern-2pt $\del$}}}

\def\msb{{\bar{\ssstyle M \kern -1pt S}}}

\def\BB0bar{B^0 {\overline B}^0}
\def\BB0dbar{B_d^0 {\overline B}_d^0}
\def\BB0sbar{B_s^0 {\overline B}_s^0}

\RequirePackage{xspace}

\usepackage{relsize}
\def\babar{\mbox{\slshape B\kern-0.1em{\smaller A}\kern-0.1em
    B\kern-0.1em{\smaller A\kern-0.2em R}}}

\def\Kbar  {\kern 0.2em\overline{\kern -0.2em K}{}\xspace}

\def\Kz    {\ensuremath{K^0}\xspace}
\def\Kzb   {\ensuremath{\Kbar^0}\xspace}
\def\KzKzb {\ensuremath{\Kz \kern -0.16em \Kzb}\xspace}
\def\Kp    {\ensuremath{K^+}\xspace}
\def\Km    {\ensuremath{K^-}\xspace}

\def\KpKm  {\ensuremath{\Kp \kern -0.16em \Km}\xspace}

\def\Dbar    {\kern 0.2em\overline{\kern -0.2em D}{}\xspace}

\def\Dz      {\ensuremath{D^0}\xspace}
\def\Dzb     {\ensuremath{\Dbar^0}\xspace}
\def\DzDzb   {\ensuremath{\Dz {\kern -0.16em \Dzb}}\xspace}
\def\Dp      {\ensuremath{D^+}\xspace}
\def\Dm      {\ensuremath{D^-}\xspace}

\def\DpDm    {\ensuremath{\Dp {\kern -0.16em \Dm}}\xspace}

\def\Bbar    {\kern 0.18em\overline{\kern -0.18em B}{}\xspace}

\def\BB      {\ensuremath{B\Bbar}\xspace} 
\def\Bz      {\ensuremath{B^0}\xspace}
\def\Bzb     {\ensuremath{\Bbar^0}\xspace}
\def\BzBzb   {\ensuremath{\Bz {\kern -0.16em \Bzb}}\xspace}
\def\Bu      {\ensuremath{B^+}\xspace}
\def\Bub     {\ensuremath{B^-}\xspace}

\def\BpBm    {\ensuremath{\Bu {\kern -0.16em \Bub}}\xspace}

\mathchardef\Upsilon="7107
\def\Y#1S{\ensuremath{\Upsilon{(#1S)}}\xspace}

\mathchardef\Deltares="7101
\mathchardef\Xi="7104
\mathchardef\Lambda="7103
\mathchardef\Sigma="7106
\mathchardef\Omega="710A

\def\Deltabar{\kern 0.25em\overline{\kern -0.25em \Deltares}{}\xspace}
\def\Lbar{\kern 0.2em\overline{\kern -0.2em\Lambda\kern 0.05em}\kern-0.05em{}\xspace}
\def\Sigbar{\kern 0.2em\overline{\kern -0.2em \Sigma}{}\xspace}
\def\Xibar{\kern 0.2em\overline{\kern -0.2em \Xi}{}\xspace}
\def\Obar{\kern 0.2em\overline{\kern -0.2em \Omega}{}\xspace}
\def\Nbar{\kern 0.2em\overline{\kern -0.2em N}{}\xspace}
\def\Xb{\kern 0.2em\overline{\kern -0.2em X}{}\xspace}

\newcommand{\tev}{\ensuremath{\mathrm{\,Te\kern -0.1em V}}\xspace}
\newcommand{\gev}{\ensuremath{\mathrm{\,Ge\kern -0.1em V}}\xspace}
\newcommand{\mev}{\ensuremath{\mathrm{\,Me\kern -0.1em V}}\xspace}
\newcommand{\kev}{\ensuremath{\mathrm{\,ke\kern -0.1em V}}\xspace}
\newcommand{\ev}{\ensuremath{\mathrm{\,e\kern -0.1em V}}\xspace}
\newcommand{\gevc}{\ensuremath{{\mathrm{\,Ge\kern -0.1em V\!/}c}}\xspace}
\newcommand{\mevc}{\ensuremath{{\mathrm{\,Me\kern -0.1em V\!/}c}}\xspace}
\newcommand{\gevcc}{\ensuremath{{\mathrm{\,Ge\kern -0.1em V\!/}c^2}}\xspace}
\newcommand{\mevcc}{\ensuremath{{\mathrm{\,Me\kern -0.1em V\!/}c^2}}\xspace}

\def\mus  {\ensuremath{\rm \,\mus}\xspace}

\def\mus        {\ensuremath{\,\mu{\rm s}}\xspace}    

\def\to                 {\ensuremath{\rightarrow}\xspace}

\def\pep2{PEP-II}

\def\gsim{{~\raise.15em\hbox{$>$}\kern-.85em
          \lower.35em\hbox{$\sim$}~}\xspace}
\def\lsim{{~\raise.15em\hbox{$<$}\kern-.85em
          \lower.35em\hbox{$\sim$}~}\xspace}

\xspace

\def\jetset74   {\mbox{\tt Jetset \hspace{-0.5em}7.\hspace{-0.2em}4}\xspace}

\begin{document}
\Title{Decay Rate Difference in the Neutral B-System: \\
       $\Delta \Gamma_{B_s}$ and $\Delta \Gamma_{B_d}$}
\bigskip
\label{ALenzStart}

%
\author{ Alexander Lenz \index{Lenz, A.} }

%
\address{Fakult{\"a}t f{\"u}r Physik\\
Universit{\"a}t Regensburg\\
D-93040 Regensburg, Germany\\
}

\makeauthor\abstracts{
We review the theoretical status of the predictions for the decay rate
differences in the neutral B-system. We find 
$(\Delta\Gamma/\Gamma)_{B_s}= (12 \pm 5) \cdot 10^{-2}$ and $(\Delta\Gamma/\Gamma)_{B_d} = (3 \pm 1.2) \cdot 10^{-3} $.}

\section{Introduction}
Recently the width difference $(\Delta\Gamma/\Gamma)_{B_s}$ of the 
$B_s$ meson CP eigenstates was measured at the Tevatron by the CDF 
Collaboration \cite{ALenz-CDF}:
\begin{equation}
\left( \frac{\Delta \Gamma}{\Gamma} \right)_{B_s} = 
0.65 ^{+0.25}_{-0.33} \pm 0.01 \; .
\end{equation}
This result can be compared with the Particle Data Group \cite{ALenz-PDG}
value
\begin{equation}
\left( \frac{\Delta \Gamma}{\Gamma} \right)_{B_s} < 0.54 \; (95 \% \mbox{C.L.})\; .
\end{equation}
In view of this new result it seems to be appropriate to
update the theoretical numbers present in the literature, see e.g.
\cite{ALenz-previousupdates}. Phenomenological aspects of the width difference
will not be discussed in this letter, we refer the interested reader to e.g.
\cite{ALenz-pheno}.
\\
The calculation of $\Delta\Gamma_{B_s}$ is performed in the framework of the heavy quark 
expansion (HQE)\cite{ALenz-HQE}, which offers the possibility to expand 
decay rates in powers of $\Lambda_{\rm QCD}/m_b$. In the case of $(\Delta\Gamma/\Gamma)_{B_s}$, 
the leading contribution is parametrically of order 
$16\pi^2 (\Lambda_{\rm QCD}/m_b)^3$. 
\begin{equation}
\left( \frac{\Delta \Gamma }{\Gamma }\right)_{B_s} = 
\frac{\Lambda^3}{m_b^3} \left( \Gamma_3^{(0)} + \frac{\alpha_s}{4 \pi} \Gamma_3^{(1)} + \ldots\right) +
\frac{\Lambda^4}{m_b^4} \left( \Gamma_4^{(0)} + \frac{\alpha_s}{4 \pi} \Gamma_4^{(1)} + \ldots\right) + \ldots
\end{equation}
Each of this $\Gamma_i^{(j)}$ consists of perturbative Wilson coefficients
and non-perturbative matrix elements.
The LO-result $\Gamma_3^{(0)}$ ($1/m_b^3$ in the HQE,
$\alpha_s^0$ in QCD and vacuum insertion approximation (VIA) for the matrix elements) 
was already calculated long time ago \cite{ALenz-LO}.
Corrections of order $1/m_b$ ($ \Gamma_4^{(0)}$) were calculated in \cite{ALenz-BBD1}
and turned out to be unexpectedly large.
Therefore terms of order $1/m_b^2$ ($ \Gamma_5^{(0)}$) should be determined 
\cite{ALenz-1m2} in order 
to check the convergence of the HQE.
${\cal O}(\alpha_s)$ radiative corrections ($ \Gamma_3^{(1)}$) were first calculated in
\cite{ALenz-BBGLN1} and confirmed in \cite{ALenz-rome03}.
The non-perturbative matrix elements of local four-quark operators 
(which appear in $\Gamma_3$) between $B$-meson states have been determined 
within the framework of QCD sum rules \cite{ALenz-Bsumrule} and lattice QCD 
\cite{ALenz-Blattice1},\cite{ALenz-Blattice2}.
The overall normalization of these matrix elements is given by the
$B_s$ decay constant, $f_{B_s}$. Since $\Delta \Gamma_{B_s}$ is proportional to $ f_{B_s}^2$
already minor changes in the numerical value of the decay constant 
have a big impact on the final prediction for $\Delta \Gamma_{B_s}$. 
Recently unquenchend lattice calculations were performed, 
which yielded large values for $f_{B_s}$ \cite{ALenz-fBlattice}. These 
numbers are in perfect agreement with recent sum rule determinations, see 
\cite{ALenz-fBsumrule} and references therein.
We will use the value which was quoted in LATTICE 2004
\cite{ALenz-fBwingate}
\begin{equation}
f_{B_s} = 245 \pm 30 \, \mbox{MeV} \, .
\end{equation}
The calculation of the next-to-leading order QCD radiative corrections
to the Wilson coefficient functions for $\Delta\Gamma_{B_s}$ was a
very important step in gaining a relieable theoretical prediction. 
First the renormalization scale dependence will be reduced compared to
the leading order prediction - unfortunateley it turned out that for
$\Delta \Gamma_{B_s}$ the remaining scale dependence is still quite large.
Second, the inclusion of ${\cal O}(\alpha_s)$ corrections 
is necessary for a satisfactory matching of the Wilson coefficients 
to the matrix elements.
The unphysical renormalization scheme dependence has to cancel between
the Wilson coefficients and the matrix elements. Since in the Wilson
coefficients this scheme dependence arises first at NLO one has to go 
beyond LO in order to obtain reliable predictions.
Moreover, the consideration of subleading QCD radiative effects was 
of conceptual interest for the construction of the HQE, 
since one could show hereby explicitly the infrared safety of the 
HQE in that order. For powerlike IR divergencies the cancellation
was already shown in \cite{ALenz-IR}.
\\
The result in \cite{ALenz-BBGLN1} was the first complete 
calculation of perturbative 
QCD effects beyond the leading logarithmic approximation to 
spectator effects in the HQE for heavy hadron decays. 
Currently NLO-QCD corrections to spectator effects are known for
the lifetime ratios of heavy hadrons 
\cite{ALenz-KN, ALenz-BBGLN2, ALenz-rome02}
and for $\Delta \Gamma_{B_d}$ and the semileptonic CP-asymmetries
\cite{ALenz-BBLN, ALenz-rome03}.
\section{Theoretical prediction of $\Delta\Gamma_{B_s}$}
\subsection{Preliminaries}
The nature of the weak interaction leads to the fact that the physical 
eigenstates of the neutral B mesons are linear combinations of the 
flavor eigenstates
\begin{eqnarray}
B_H & := & p \; B + q \; \bar{B} \, ,
\\
B_L & := & p \; B - q \; \bar{B} \, .
\end{eqnarray}
Three measurable quantities can be deduced from this particle-antiparticle
mixing:
\begin{eqnarray}
\Delta M & :=& M_H - M_L \, ,
\\
\Delta \Gamma & :=& \Gamma_L - \Gamma_H  \, ,
\\
a_{fs} & = & - 2  \left( \left| \frac{q}{p}\right| - 1 \right) \, .
\end{eqnarray}
$a_{fs}$ describes  CP asymmetries in flavor specific B decays, which are often 
called semi-leptonic CP asymmetries. This quantity is discussed e.g. in
\cite{ALenz-BBLN, ALenz-rome03}. In the following we restrict ourselves
to $\Delta \Gamma$.
The decay rate difference can be expressed as the matrix element of
the transition operator ${\cal T}$
\begin{equation}
\Delta \Gamma = - \frac{1}{m_{B_s}} \langle \overline{B}_s | {\cal T} | B_s \rangle \, ,
\end{equation}
which consists of a double insertion of the {$ \Delta B=1$} effective Hamiltonian
\begin{equation}
{\cal T}= \mbox{Im} \; i \int d^4x 
 T \left[ {\cal H}_{eff}(x), {\cal H}_{eff} (0) \right] \, .
\end{equation}
Formally one performs now a operator product expansion for the transition
operator, graphically one matches the
{$ \Delta B=1$} double insertion to a {$  \Delta B=2$} insertion.
\subsection{Leading order}
\begin{figure}[htb]
\begin{center}
\epsfig{file=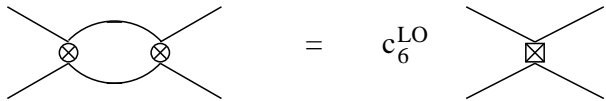,height=80mm, angle=270}
\caption{LO matching condition for $\Delta \Gamma$.}
\label{fig:ALenz-matchlo}
\end{center}
\end{figure}

In LO in the HQE the matching equation is described by fig.
(\ref{fig:ALenz-matchlo}). The l.h.s. of fig.(\ref{fig:ALenz-matchlo})
corresponds to the double insertion of the effective $\Delta B = 1$
Hamiltoninan. By calculating this loop diagram one obtains the r.h.s.,
which consists of the Wilsoncoefficent $c_6^{LO}$ and a four-quark $\Delta B = 2$ 
operator.
One can express the transition operator in the following form
\begin{equation}
{\cal T}  =  -\frac{G^2_F m^2_b}{12\pi}(V^*_{cb}V_{cs})^2
\, \left[ F(z) Q(\mu_2)+ F_S(z) Q_S(\mu_2) \right]
\end{equation}
with the Wilson coefficients $F$ and $F_S$ ($z = m_c^2/m_b^2$) and the 
following $\Delta B = 2$-operators 
\begin{eqnarray}
Q & = & (\bar{b}_i s_i)_{V-A} \cdot (\bar{b}_j s_j)_{V-A}
\\
Q_S & = & (\bar{b}_i s_i)_{S-P} \cdot (\bar{b}_j s_j)_{S-P}
\end{eqnarray}
The color-rearranged operators which arise during the calculation have been
eliminated via
\begin{eqnarray}
\tilde Q & = & Q 
\\
\tilde Q_S & = & - Q_S - \frac{1}{2} Q + {\cal O} (\alpha_s) 
+ {\cal O} \left(\frac{1}{m_b} \right) 
\label{eq:ALenz-tilde}
\end{eqnarray}
The matrix elements of $Q$ and $Q_S$ can be parametrized in terms of
the decay constant $f_{B_s}$ and bag parameters $B$ and $B_S$.
\begin{eqnarray}
\langle \bar{B}_s | Q | B_s \rangle & = &
\frac{8}{3} {f^2_{B_S}} M^2_{B_S} {B}
\\
\langle \bar{B}_s | Q_S | B_s \rangle & = &
- \frac{5}{3} {f^2_{B_S}} M^2_{B_S} 
\frac{M^2_{B_S}}{(\bar{m}_b + \bar{m}_s)^2}
{B_S}
\end{eqnarray}
Assuming VIA for the matrix elements, 
which corresponds to setting the bag parameters equal to one, we get
with $z = 0.085, \bar m_b = 4.2 $ GeV, $\bar m_s = 0.1$ GeV, 
$V_{cb} = 40.1 \cdot 10^{-3}, m_b = 4.8$ GeV
\begin{equation}
\left( \frac{ \Delta \Gamma}{\Gamma} \right)_{B_{s}} = 
{\cal O} (30)  \% \, .
\end{equation}
\subsection{Next-to-Leading QCD corrections}
If one dresses the diagrams of fig. (\ref{fig:ALenz-matchlo}) with 
one gluon in all possible ways one gets the NLO QCD correction to 
the Wilson coefficients.
At this level one has to take the $\alpha_s$-corrections 
to eq. (\ref{eq:ALenz-tilde})
into account.
The NLO-QCD calculation was performed in 
\cite{ALenz-BBGLN1} and \cite{ALenz-rome03} and gives a sizeable reduction of 
the LO result:
\begin{equation}
\left( \frac{ \Delta \Gamma}{\Gamma} \right)_{B_{s}} = 
{\cal O} (24)  \% \, .
\end{equation}
Unfortunateley it turned out that the reduction of the renormalization scale
dependence was not very pronounced. An improvement of this point would 
require the calculation of $\alpha_s^2$-corrections, which will be a hard
endeavour,  even though the three-loop anomalous dimensions of 
the effective Hamiltonian are now known \cite{ALenz-3loop}.
\subsection{Lattice evaluation of the matrix elements}
Since the scheme dependence is now visible due to the NLO QCD calculation,
a next step to obtain a relieable prediction for $\Delta \Gamma_{B_s}$ 
is the inclusion of lattice predictions for the bag parameters instead of VIA.
Now the unphysical scheme dependence, which can be numerically large cancels 
up to effects of order $\alpha_s^2$. 
With \cite{ALenz-Blattice1}
\begin{eqnarray}
B & = & 0.87 \pm 0.06
\\
B_S & = & 0.84 \pm 0.05
\end{eqnarray}
one obtains again a reduction of the final number:
\begin{equation}
\left( \frac{ \Delta \Gamma}{\Gamma} \right)_{B_{s}} = 
{\cal O} (20)  \% \; .
\end{equation}
\subsection{Power corrections}
Till now we were setting the momentum of the light quark in the B-meson 
to zero. Power corrections ($ \equiv 1/m_b$-corrections) can be obtained by expanding
the transition operator in powers of the light quark momentum. In addition
one has to take the $1/m_b$-corrections to eq. (\ref{eq:ALenz-tilde})
into account. The calculation of the $1/m_b$-corrections was performed first 
in \cite{ALenz-BBD1}.
In this order of the HQE operators of dimension 7 appear. Some of these
operators can be rewritten with the help of e.o.m. to dimension 6
operators, the remaining operators have to be estimated by VIA.
Once again we get a sizeable reduction of our prediction:
\begin{equation}
\left( \frac{ \Delta \Gamma}{\Gamma} \right)_{B_{s}} = 
{\cal O} (12)  \% \, .
\end{equation}
The power corrections turn out to be the most important corrections and 
at the same time the least well known ones.
In order to improve our knowledge about this corrections several
tasks have to be completed:
\begin{itemize}
\item \underline{Test the HQE expansion:} the calculation of 
      $1/m^2$-corrections is under way \cite{ALenz-1m2}. 
\item \underline{Matrix elements of dimension 7 operators:}
      As was noted already in \cite{ALenz-rome03} some of these power 
      suppressed operators can be obtained from the lattice evaluation 
      in \cite{ALenz-Blattice1}. Despite this progress it is still 
      necessary to have a relieable determination of the 
      remaining dimension 7 operators.
\item \underline{QCD corrections to power corrections:} with a lattice
      determination of the dimension 7 operators at hand it might be 
      worthwhile to calculate $\Gamma_4^{(1)}$.
\end{itemize}
\subsection{The Final number}
We have here the very special situation that all corrections have a 
negative sign and are quite sizeable. Moreover we have an additional source 
of uncertainty.
So far we were actually only calculating $\Delta \Gamma_{B_s}$, the ratio {$(\Delta \Gamma/\Gamma)_{B_s}$}
can be obtained in different ways (A, B and C)
\large
\begin{equation}
\begin{array}{cclcl}
\left(\frac{\Delta\Gamma}{\Gamma}\right)_{B_s}^A  & = & \Delta\Gamma_{B_s}   \tau_{B_{s/d}}
& = &
\frac{G_F^2}{12 \pi}  m_b^2 V_{cb}^2 \tau_{B_{s/d}} { f_{B_s}^2} K \, ,
\\ \\
\left(\frac{\Delta\Gamma}{\Gamma}\right)_{B_s}^B 
& = &
\Delta\Gamma_{B_s} \frac{1}{\Gamma_{sl}}  B_{sl} \frac{\tau_{B_s}}{\tau_{B_d}}
& = &
16\pi^2 \frac{B_{sl}}{g(z)\eta_{sl} m^3_b} \frac{\tau_{B_s}}{\tau_{B_d}} f^2_{B_s} K \, ,
\\ \\
\left(\frac{\Delta\Gamma}{\Gamma}\right)_{B_s}^C 
& = &
\frac{\Delta\Gamma_{B_s}} {\Delta M_{B_s}}  \frac{\Delta M_{B_s}}{\Delta M_{B_d}} \Delta M_{B_d} \tau_{B_s}
& = &
\frac{\pi}{2 M_W^2} \frac{\Delta M_{B_d}}{M_{B_d}} \frac{m_b^2 V_{cb}^2 { \xi^2} \tau_{B_s}} {(V_{tb} { V_{td}})^2 \eta_B S_0(x_t) B} K \, ,
\end{array}
\end{equation}
\normalsize
with
\begin{displaymath}
K =  M_{B_s}  V_{cs}^2 
\left[ F \langle Q \rangle + F_S \langle Q_S \rangle \right]\, .
\end{displaymath}
Unfortunateley we have here a similar situation like in the
case of the missing charm puzzle \cite{ALenz-nc}, that different
normalizations lead to big numerical effects.
Method C, which was used e.g. in \cite{ALenz-rome03} tends to give
values which are about $25\%$ smaller than method B,  which was used 
e.g. in \cite{ALenz-BBGLN1}. In this letter we were using method B, for 
future estimates we suggest to use method A, see \cite{ALenz-1m2}. 
\\
Putting everything together and estimating the dominant errors we get
\begin{eqnarray}
\left( \frac{ \Delta \Gamma}{\Gamma} \right)_{B_{s}} & = &
 \left( \frac{f_{B_s}}{245 \mbox{MeV}} \right)^2
 \left[ 0.234 B_S (m_b) - 0.086 + 0.008 B(m_b) \right]
\\
&&
\nonumber \\
& = &
\left( 12 \pm 5 \right)  \% \; .
\end{eqnarray}
\section{Theoretical prediction of $\Delta \Gamma_{B_d}$}
In principle the calculation of $\Delta \Gamma_{B_d}$ proceeds in the same way,
but one has to keep in mind that in this case different CKM structures 
contribute with a similar strength (order $\lambda^6$ in the Wolfenstein parameter 
$\lambda$), while in the case of $\Delta \Gamma_{B_s}$ the contribution of two internal
charm quarks is leading by two powers of $\lambda$. 
\begin{equation}
\begin{array}{|c|c|c|}
\hline
\mbox{internal quarks} & \; \; \Delta \Gamma_{B_s}   \; \;&  \; \;\Delta \Gamma_{B_d} \; \;
\\
\hline
uu  & \lambda^8 & \lambda^6
\\
\hline
uc  & \lambda^6 & \lambda^6   
\\ 
\hline
cu  & \lambda^6 & \lambda^6   
\\
\hline
cc  & \lambda^4 & \lambda^6   
\\
\hline
\end{array}
\end{equation}
The $uu$ and the $cc$ contribution can be taken from the $\Delta \Gamma_{B_s}$-calculation,
while the $uc$ and $cu$ contributions have to calculated anew.
$\Gamma_3^{(0)}$ was calculated in \cite{ALenz-LO}, $\Gamma_4^{(0)}$ was calculated in 
\cite{ALenz-CSKIM} and $\Gamma_3^{(1)}$ was calculated in \cite{ALenz-BBLN} and
\cite{ALenz-rome03}.
\begin{eqnarray}
\left( \frac{ \Delta \Gamma}{\Gamma} \right)_{B_{d}} & = &
\left( 3 \pm 1.2 \right)  \cdot 10^{-3} \; .
\end{eqnarray}
\section{Outlook for $\Delta \Gamma_{B_s}$}
What do we expect for the future?
First we are waiting eagerly for the D0 number for $\Delta \Gamma_{B_s}$  and we of 
course expect much smaller errors in the future.
From the theory side we have the following what-to -do-list
\begin{itemize}
\item Calculation of $1/m_b^2$-corrections: $\Gamma_5^{(0)}$ 
\item Lattice determination of dimension 7 operators for $\Gamma_4$
\item Calculation of $\alpha_s$-corrections to the $1/m_b$corrections: $\Gamma_4{(1)}$
\item Calculation of $\alpha_s^2$-corrections to the leading term: $\Gamma_3{(2)}$
\end{itemize}
It was shown in \cite{ALenz-GRO} that new physics effects can not enhance 
$\Delta\Gamma_{B_s}$ compared to the standard model value.
If after all these efforts the central experimental und theoretical 
numbers stay at their current values, this would probably be a signal 
of local quark-hadron duality violation. The operator product expansion 
of the transition operator is based on the duality assumption.
Little is known in QCD about the actual numerical size of duality-violating 
effects. Experimentally no violation of local quark-hadron duality
in inclusive observables of the $B$-meson sector has 
been established so far. Comparison of experiment and theory for 
$\tau (B^+)/\tau (B_d)$ supports the duality assumption, but in that case we have 
only one heavy charm in the intermediate state, compared to two charm quarks 
in the $\Delta  \Gamma_{B_s}$-case. 
In \cite{ALenz-ALE93} it has been shown that for $\Delta\Gamma_{B_s}$
local duality holds exactly in the simultaneous limits of small
velocity ($\Lambda_{QCD}\ll$ $m_b-2m_c\ll m_b$) and large number of
colours ($N_c\to\infty$). In this case
\begin{equation}\label{ALenz-dglim}
\left(\frac{\Delta\Gamma}{\Gamma}\right)_{B_s}=
\frac{G^2_F m^3_b f^2_{B_s}}{4\pi}|V_{cs}V_{cb}|^2\,
\sqrt{2-4\frac{m_c}{m_b}}\, \tau_{B_s}\approx 0.18.
\end{equation}
It is interesting that the numerical value implied
by the limiting formula (\ref{ALenz-dglim}) appears to be quite realistic.
\section*{Acknowledgments}
I would like to thank the organizers of FPCP2004 for the invitation and the 
financial support, M. Beneke, G. Buchalla, C. Greub and U. Nierste for 
the pleasant collaboration
and Fermilab and DFG for financial support, while calculating the 
$1/m^2$-corrections.

%
\label{ALenzEnd}

\end{document}